\documentclass{CHEP2006}
\usepackage{graphicx}

\begin{document}
\title{HZTOOL AND RIVET: TOOLKIT AND FRAMEWORK FOR THE COMPARISON OF
  SIMULATED FINAL STATES AND DATA AT COLLIDERS}

\author{
  B. M. Waugh\thanks{waugh@hep.ucl.ac.uk}, University College London, UK\\
  H. Jung, DESY, Hamburg, Germany\\
  A. Buckley, Durham University, Durham, UK\\
  L. L\"onnblad, Lund University, Lund, Sweden\\
  J. M. Butterworth and E. Nurse, University College London, UK
}

\maketitle

\begin{abstract}
A common problem in particle physics is the requirement to reproduce
comparisons between data and theory when the theory is a (general
purpose) Monte Carlo simulation and the data are measurements of final
state observables in high energy collisions.  The complexity of the
experiments, the obervables and the models all contribute to making
this a highly non-trivial task.

We describe an existing library of Fortran routines, HZTool, which
enables, for each measurement of interest, a comparable prediction to
be produced from any given Monte Carlo generator. The HZTool library
is being maintained by CEDAR, with subroutines for various
measurements contributed by a number of authors within and outside the
CEDAR collaboration.

We also describe the outline design and current status of a
replacement for HZTool, to be called Rivet (Robust Independent
Validation of Experiment and Theory). This will use an object-oriented
design, implemented in C++, together with standard interfaces (such as
HepMC and AIDA) to make the new framework more flexible and extensible
than the Fortran HZTool.
\end{abstract}

\section{INTRODUCTION}

\begin{figure}[htb]
\centering
\includegraphics*[width=65mm]{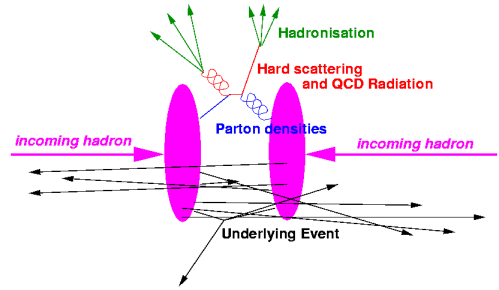}
\caption{A hadronic collision.}
\label{collision}
\end{figure}

While the Standard Model provides methods of making some calculations
to an impressive degree of precision, there are still many areas where
first-principles calculations must be supplemented by a more
phenomenological approach. This is particularly the case in collisions
involving hadrons, where the non-perturbative regime of QCD plays a
major role, especially in the hadronisation stage, as do parton
density functions, fragmentation and -- especially at high energies --
the underlying event, i.e.\ the part of the hadronic final state that
is not coherent with the hard scatter.
These effects are illustrated in figure~\ref{collision}.

Calculations using only the well understood ``hard'' regime of QCD can
provide only an incomplete test of the Standard Model and of theories
that go beyond it. Models that include these more complicated effects
and provide a full description of the final state, generally
implemented as Monte Carlo event generators, fill a number of
roles in high-energy physics. As well as providing
predictions that can be used to test the agreement between theory and
experiment, they are extensively used to correct for detector effects
in making measurements, to estimate the background to processes of
interest, and to make predictions of both standard and exotic
processes for use in planning analyses and designing future
accelerators and detectors.

The models used to describe these processes typically have a large
number of free or weakly constrained parameters, which must be tuned
to the data. This is a complicated task, as the same parameter may
affect the predictions for different measurements in widely differing
way. It is relatively easy to tune a parameter or limited set of
parameters in order to provide a good description of a particular
observable, only to spoil the agreement of the model with other
data. It is therefore important to compare models simultaneously to a
wide range of data from different colliding particles in different
energy regimes, including older data from experiments that may no
longer be running, but still constrain the range of acceptable models.

It is to provide a solution to this problem that HZTool~\cite{hztool} has been
developed, and Rivet~\cite{rivet} is now being created. These packages form an
essential part of the wider programme of work being carried out by the
CEDAR~\cite{cedar} collaboration.

\section{REQUIREMENTS}

In order to enable meaningful comparisons between data and Monte Carlo
predictions, both HZTool and Rivet must incorporate routines or modules that
encapsulate information about the meaning of the 
measured quantities that goes beyond the simple numerical values
stored in the HEPDATA~\cite{hepdata} database.
These routines must make the data
available for future studies and model comparisons that may not have
been possible when the data was published. The precise meaning of an
observable and the procedure used to produce it may be hard to
reconstruct from the published paper. It is therefore useful for the routine to
be written, if possible, by the authors of the original analysis.

The framework must cleanly separate the analysis routine from any
individual event generator, reading the event information only via a common
interface. This means that the routine can be used unchanged with any
generator, rather than being restricted to those in use at the time
the measurement was made and published.

HZTool and Rivet provide essential low-level tools enabling the
comparison of theoretical predictions with experimental data.
They may be used as stand-alone packages, but there
are also a number of ways in which their functionality can be extended.
JetWeb~\cite{jetweb} uses
HZTool as a back end (and will in future use Rivet) but adds a web
interface and a database of predictions, and is described in another
contribution to these proceedings~\cite{hepdata-jetweb}.
This reduces duplication of effort by enabling predictions generated
by one user to be reused by others,
but still requires the user to specify the model
of interest including any parameter values that differ from the
defaults. There is clearly scope for higher-level tools that
automatically vary parameters in order to find improved
models. Professor~\cite{professor} is an example of an attempt to do
this.

\section{HZTOOL}

The HZTool framework,
written in Fortran 77, was originally developed
as part of the workshop on Future Physics at HERA in 1997. It started
with an emphasis on HERA results, but subsequently it has been
expanded to cover a wider range of measurements. HZTool has recently
become part of the CEDAR project, and some further
development has taken place in order to improve its modularity and to
bring together some of the different versions that were in use into a
single easily available package. Development has been moved to the
HepForge~\cite{hepforge} environment.

Each included experimental analysis has a corresponding Fortran
subroutine in HZTool. This subroutine
takes events from a Monte Carlo event generation run and fills
histograms in order to produce data and plots that can be compared
directly with the experimental results.
In the most recent version, HZTool is purely a library of analysis
routines and associated utilities such as jet finders, which may be
used by a number of different routines.
The main programs and steering routines that were
formerly part of the package have been split off into a separate
package, HZSteer, which is designed chiefly as an interface between
JetWeb and HZTool, although it can also be used to run
HZTool independently of JetWeb. This separation means that HZTool is
completely independent of any Monte Carlo generators, with all such
dependencies existing only in HZSteer.

The HEPEVT common block is used in order to make the code independent
of the particular generator used. Histogramming is done using the
HBOOK histogramming package that forms part of the
CERNLIB~\cite{cernlib} library.

As part of the CEDAR project, HZTool is being adapted to provide model
descriptions in HepML~\cite{hepml,hepdata-jetweb} format and to use
HEPDATA as a source of experimental data rather than duplicating the
information in Fortran DATA statements.

\begin{figure}[htb]
\centering
\includegraphics*[width=80mm]{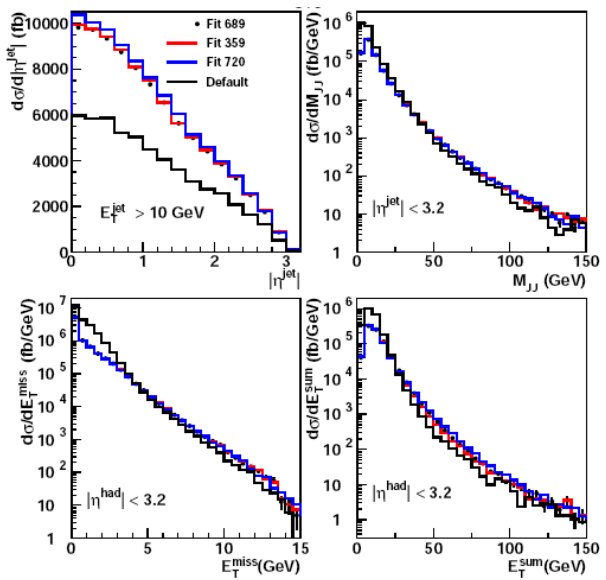}
\caption{Predictions of various HERWIG models for jet production in
  $e^+e^-$ collisions at a centre-of-mass energy of
  500~GeV\@. From~\cite{lincoltuning}.}
\label{lincol}
\end{figure}

Figure~\ref{lincol} shows some histograms created using HZTool as part
of a study of Monte Carlo tuning for the International Linear
Collider~\cite{photoproduction,lincoltuning}.
The continuous black line shows the
predictions of the default HERWIG~\cite{herwig} model for dijet production in
$\gamma\gamma$ events in $e^+e^-$ collisions at a centre-of-mass energy
of 500~GeV\@. The other curves show the predictions of models using
HERWIG with parameters tuned to a range of jet production data from
HERA, LEP and the Tevatron. The results show the benefit of systematic
tuning, since the models that have been tuned to a wide range of data
show reasonable agreement with one another, despite differing
significantly from the generator default settings.

\section{RIVET}

Rivet uses an object-oriented design, implemented in C++, to make the
framework easier to maintain than HZTool, as well as making it easier
to interact with the more recent C++ Monte Carlo programs.

Wherever possible, interaction between Rivet and other software and
data will use common interfaces rather than introducing unnecesary
dependences on particular packages.
This includes the use of HepMC~\cite{hepmc} for accessing information
about the generated
events. It is important that when a new generator (or version) is
released, its predictions can be compared to all data in the Rivet
library without the need to modify each analysis routine.
The use of HEPDATA for access to data and the HepML interface for
input and output of Monte Carlo configurations is being incorporated in
Rivet from the start.
The creation of histograms will be dealt with using the
AIDA~\cite{aida} interface.

Rather than reimplement jet finders and other general analysis
utilities already available elsewhere, Rivet will make use of existing
libraries such as KtJet~\cite{ktjet}. Where suitable general-purpose
(i.e.\ independent of any particular experiment's framework) tools are
not available, we will try where possible to make our solutions
available and usable without depending on Rivet.

As in the case of HZTool and HZSteer, the steering and all
generator-dependent code is in a separate package, RivetGun, so that
the Rivet library can be compiled without any generator code and does
not need to be recompiled in order to use it with a new generator.

Each physics paper implemented in Rivet will be represented by an
analysis ``module''.
The analysis modules have access to a collection of {\tt Projection}
classes, each of which calculates some property of an event. A
projection may be built up from a series of simpler projections. For
example, a projection that finds jets in an event might call on
another projection to select particles in the event.

Rivet will run roughly as follows:
\begin{itemize}
\item A run will in general use a single generator with one
  combination of beam particles, energies and steering
  parameters.
\item A {\tt RivetHandler} will be instantiated, and will be given a
  collection of {\tt Analysis} objects to use, each corresponding to a
  physics paper with measurements to be compared to the generator
  predictions.
\item {\tt Analysis} classes will inherit from a common base class,
  {\tt AnalysisBase}. Each such class will use {\tt Projection} objects
  to calculate event properties and implement kinematic cuts.
\item A {\tt Projection} takes an {\tt Event} and constructs some
  property of the event, which may be a single number (e.g. $Q^2$) or
  a more complex object such as a set of jets boosted to a particular
  coordinate frame.
\item Each {\tt Projection} object is instantiated along with the {\tt
  Analysis} at the beginning of a run, but only acts on the current
  event if and when it is actually asked for the information. Once it
  has carried out its calculation it caches the result, which can then
  be accessed again by the same {\tt Analysis} or a different {\tt
  Analysis}, thus saving a significant amount of time in the case of
  e.g. a jet algorithm.
\item Once an {\tt Analysis} has used a series of {\tt Projection}s to
  calculate the quantities of interest and apply any necessary cuts,
  it uses the Rivet histogramming code to fill the relevant
  histograms. This will probably involve simply using the AIDA
  interfaces to invoke whatever histogramming package is chosen by the
  user.
\end{itemize}

\section{CONCLUSIONS}

HZTool has been actively used for eight years, both as a stand-alone
application and as the back-end to the JetWeb interface. It will
continue to be maintained as long as the knowledge encoded in its
routines is not available in another form. However, active development
will soon cease in favour of Rivet, and once Rivet is available
collaborations will be encouraged to provide Rivet modules instead of
HZTool subroutines.

These frameworks provide a fundamental part of the CEDAR programme,
making use of other CEDAR projects such as HEPDATA, and providing an
input to higher-level tools such as JetWeb, and potentially to more
automated tuning systems such as Professor.

While a clear timescale is set by LHC start-up, and the short-term
goals of CEDAR are directed towards this, the project will not stop
there but will continue to provide a valuable resource for the Linear
Collider and other future facillities.

\section{ACKNOWLEDGEMENTS}

The CEDAR team would like to thank the UK Particle Physics and
Astronomy Research Council (PPARC) for their generous support of
CEDAR.


\end{document}